\documentclass[amsmath,amssymb,aps,pra,reprint,superscriptaddress,footnoteinbib,longbibliography]{revtex4-1}
\usepackage{amsmath,amssymb}
\usepackage[english]{babel}
\usepackage{ bbold }
\usepackage{upgreek}
\usepackage{dsfont}
\usepackage{graphicx}
\usepackage{comment}
\usepackage{braket}
\usepackage{colortbl}
\usepackage{blindtext}
\usepackage{xr-hyper}
\usepackage[hidelinks]{hyperref}
\allowdisplaybreaks

\begin{document}

\title{Force-gradient sensing and entanglement via feedback cooling of interacting nanoparticles}

\author{Henning Rudolph}
\affiliation{University of Duisburg-Essen, Faculty of Physics, Lotharstra\ss e 1, 47057 Duisburg, Germany}

\author{Uro\v{s} Deli\'c}
\affiliation{University of Vienna, Faculty of Physics, Boltzmanngasse 5, A-1090 Vienna, Austria}

\author{Markus Aspelmeyer}
\affiliation{University of Vienna, Faculty of Physics, Boltzmanngasse 5, A-1090 Vienna, Austria}
\affiliation{Austrian Academy of Sciences, Institute for Quantum Optics and Quantum Information (IQOQI) Vienna, Boltzmanngasse 3, A-1090 Vienna, Austria}

\author{Klaus Hornberger}
\affiliation{University of Duisburg-Essen, Faculty of Physics, Lotharstra\ss e 1, 47057 Duisburg, Germany}

\author{Benjamin A. Stickler}
\affiliation{University of Duisburg-Essen, Faculty of Physics, Lotharstra\ss e 1, 47057 Duisburg, Germany}

\begin{abstract}
We show theoretically that feedback-cooling of two levitated, interacting nanoparticles enables differential sensing of forces and the observation of stationary entanglement. The feedback drives the two particles into a stationary, non-thermal state which is susceptible to inhomogeneous force fields and which exhibits entanglement for sufficiently strong inter-particle couplings. We predict that force-gradient sensing at the zepto-Newton per micron range is feasible and that entanglement due to the Coulomb interaction between charged particles can be realistically observed in state-of-the-art setups.
\end{abstract}

\maketitle

\section{Introduction}The key to precision sensing lies in a thorough isolation from environmental perturbations. Levitating nanoparticles with lasers in ultra-high vacuum presents a promising platform, as optical fields grant precise control over the particle motion, while the environment can be efficiently shielded \cite{millen2020,gonzalez2021}. Their exquisite isolation renders these systems promising for future force and torque sensing technologies \cite{ranjit2016, hempston2017, ahn2020}, for tests of the quantum superposition principle with increasingly macroscopic objects \cite{arndt2014,millen2020part2,stickler2021}, for the observation of mechanical entanglement \cite{rudolph2020, krisnanda2020, qvarfort2020, rakhubovsky2020, chauhan2020,cosco2021,weiss2021,brandao2021}, for explorations of physics beyond the standard model \cite{moore2021,carney2021,afek2022} and for probing the quantumness of gravity \cite{bose2017,marletto2017,Chevalier2020,carney2021b,pedernales2022,streltsov2022significance}.

Levitated particles have been cooled to the motional ground state by two different methods \cite{delic2020,magrini2021,tebbenjohanns2021,ranfangni2022twodimensional,kamba2022optical}. Coherent scattering cooling uses the scattering of red-detuned tweezer light into a high finesse cavity \cite{delic2019,windey2019,gonzalez2019,rudolph2021}, efficiently reducing the effect of laser phase noise heating \cite{meyer2019} and holding the prospects of also cooling rotational degrees of freedom \cite{schafer2021}. In contrast, feedback cooling \cite{vovrosh2017, seberson2019, conangla2019, tebbenjohanns2020, dania2021, van2021} uses the information extracted from the Rayleigh-scattered tweezer light to apply feedback forces and cool the particle motion. Feedback cooling circumvents limitations posed by shot noise if the information leaking out of the system can be detected and utilized sufficiently well \cite{magrini2021,tebbenjohanns2021}. Feedback control of levitated particles has also been proposed for optimal control of their quantum state \cite{setter2018, conangla2019,cosco2021,magrini2021,weiss2021}, which is ultimately limited by the computational resources of the feedback loop.

In this work, we show theoretically how feedback techniques can drive two interacting nanoparticles into a stationary state close to the two-particle groundstate. Our calculations demonstrate that this enables sensing of local force gradients on the $\text{zepto-Newton}/\mu\text{m}$ scale and that the interaction leads to squeezing of the relative motion. We predict under which conditions the two particles generate stationary Gaussian entanglement, for instance due to Coulomb interaction, where feedback cooling replaces the cryogenic cooling on which clamped experiments typically rely \cite{ockeloen2018, riedinger2018, marinkovic2018}. Inter-particle coupling due to the Coulomb force has been recently studied experimentally with levitated nanoparticles \cite{penny2021,rieser2022}. Our scheme is feasible with state-of-the-art technology \cite{tebbenjohanns2021,magrini2021}, requiring merely that both normal modes of the two-particle motion can be measured and feedback cooled individually, as recently accomplished for non-interacting particles \cite{vijadan2022scalable}.

\begin{figure}[b]
	\centering
	\includegraphics[width=1\linewidth]{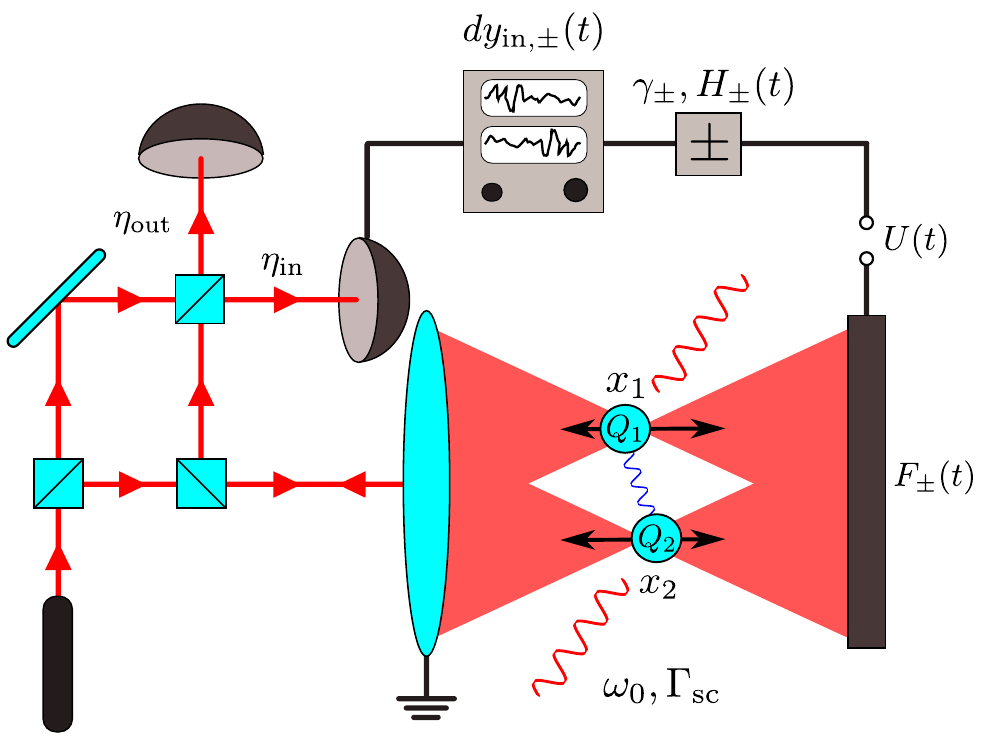}
	\caption{Two spherical, charged nanoparticles (blue spheres) are levitated in two tweezer traps with identical trapping frequencies $\omega_0$ and recoil heating rates $\Gamma_{\rm sc}$, whose collimation lens is grounded. The particles motion along the optical axis $x_{1,2}$ is detected with an efficiency $\eta_{\rm in}$ and the measurement signals of the sum and difference modes $dy_{\text{in},\pm}(t)$ are monitored. The signals are convoluted with the feedback functions $H_\pm(t)$ and multiplied with damping constants $\gamma_\pm$ to apply voltage $U(t)$, leading to feedback forces $F_\pm$. A second detector measures the particles motion with efficiency $\eta_{\rm out}$, which is not part of the feedback loop, and is additionally used to detect external forces and interparticle entanglement.}
	\label{fig:sketch}
\end{figure}

\section{Feedback master equation}The proposed experimental setup is depicted in Fig.~\ref{fig:sketch}. Two nanoparticles of mass $m$, electric susceptibility $\chi_{\rm e} = 3(\varepsilon_{\rm r} - 1)/(\varepsilon_{\rm r} + 2)$, dielectric permittivity $\varepsilon_{\rm r}$, mass density $\varrho$, and volume $V$ are trapped in two parallel tweezers with approximately equal wavenumbers $k$, Rayleigh ranges $x_{\rm R}$, and drive powers $P$. Our assumption of (almost) equal particles and tweezers, which is for simplicity and ease of notation, can be readily lifted. The tweezer foci are separated by a distance $d$ and lie in the plane perpendicular to both tweezer propagation directions. The light scattered off each particle is collected and distributed among four homodyne detectors: two in-loop detectors with efficiencies $\eta_{\rm in}$ responsible for feedback cooling the particles, and two out-of-loop detectors with efficiencies $\eta_{\rm out}$ required for independently monitoring the particle motion and for force measurements. For simplicity, we assume the tweezers to be detuned from each other to avoid interference between the tweezer outputs in the detection. Otherwise, an appropriate spectral filter would be required to access all normal modes of the mechanical system.

The particle motion perpendicular to the optical axis is assumed to be sufficiently cooled that coupling to the motion along the optical axis can be neglected \cite{magrini2021,tebbenjohanns2021}. In that case we may restrict the discussion to the coordinates along the optical axis,  denoted by $x_{1,2}$.

For harmonically trapped particles, whose diameter is much smaller than the optical wavelength, the trapping frequency is given by $\omega_0 = \sqrt{\chi_{\rm e} k P/\pi c\varrho x_{\text{R}}^3}$. Incoherent scattering of tweezer photons leads to recoil heating with the rate \cite{rudolph2021}
\begin{align}\label{rayleighrate}
    \Gamma_{\rm sc} = \omega_0 \frac{\chi_{\rm e}k^5 V x_{\text{R}}^2}{60\pi}\left[2 + 5\left(1-\frac{1}{kx_{\text{R}}}\right)^2\right].
\end{align}
The particles hold excess charges $Q_{1,2}$,  $|Q_1|\neq|Q_2|$, enabling one to apply the feedback forces $F_{1,2}(t)$ via a homogeneous electric field and thereby to cool the motion of both particles. The two particles interact via the Coulomb force, which for small displacements from the tweezer foci is given by $V_{\rm int} = m\omega_0 g (x_1-x_2)^2$. The coupling constant  can be expressed as $g = -Q_1 Q_2/8\pi m\omega_0\varepsilon_0 d^3$, so that $g>0$ for attractive interactions and $g<0$ for repulsion. In addition to this electrostatic interaction, the particles might interact via conservative optical binding forces \cite{rieser2022}, which would increase the coupling constant $g$. In Appendix \ref{equilibriumshift} we show that the electrostatic interaction also displaces the particles along their connecting axis, potentially changing the effective trapping frequency and recoil rate.

Our aim is to measure the difference between the two forces $K_{1,2}(t)$, which are acting on the two particles due to an additional inhomogeneous field. The Hamiltonian describing the particle dynamics can be expressed in terms of the sum and difference motion operators, $x_\pm = (x_2 \pm x_1)/\sqrt{2}$ and $p_\pm = (p_2 \pm p_1)/\sqrt{2}$ as
\begin{align}\label{hamiltonian}
    H = \sum_{s = \pm} \left (\frac{p_s^2}{2m} + \frac{m\omega_s^2}{2}x_s^2 - F_s(t)x_s - K_s(t)x_s \right ).
\end{align}
Here we defined the sum and difference mode frequencies $\omega_+ = \omega_0$ and $\omega_-^2 = \omega_0^2 + 4g\omega_0$ as well as the mean and offset forces $F_\pm (t) = [F_2(t) \pm F_1(t)]/\sqrt{2}$ and $K_\pm (t) = [K_2(t) \pm K_1 (t)]/\sqrt{2}$. The center-of-mass motion of the two particles thus serves to measure the mean external force while their differential motion senses the mean force gradient.

The homodyne detectors monitor the particle motion by generating the measurement outcomes $dy_{1,2}^r$, where $r\in\lbrace \text{in,out} \rbrace$ denotes the the in-loop and out-of-loop measurement channels. Given the two-particle state $\rho$ and combining the measurement outcomes to $dy_{r\pm} = (dy_2^r \pm dy_1^r)/\sqrt{2}$ yields \cite{jacobs2014}
\begin{align}\label{outcome}
    dy_{rs} = \langle x_s \rangle dt + \frac{L}{\sqrt{\eta_r}} dW_{rs}
\end{align}
where $s\in\lbrace+,-\rbrace$. Here, $\langle \cdot \rangle = {\rm tr}(\rho \,\cdot)$ denotes the quantum expectation value and $dW_{rs}$ are statistically independent Wiener processes, $\mathbb{E}[dW_{rs}]=0$ and $\mathbb{E}[dW_{rs} dW_{r's'}] = \delta_{rr'}\delta_{ss'}dt$, where $\mathbb{E}[\cdot]$ is the ensemble average over the measurement outcomes. The Wiener increments model the photon shot noise due to the local oscillators used for the homodyne detection. The parameter $L^2 = \hbar/8m\omega_0\Gamma_{\rm sc}$ describes the accuracy of the position measurements as determined by the recoil heating rate, and the quantum efficiencies $\eta_r$ quantify how much information can be extracted from the scattering fields, taking into account that the information is not uniformly distributed for all scattering directions \cite{magrini2021,tebbenjohanns2021}. Note that the setup  implies $\eta_{\rm in} + \eta_{\rm out} \leq 1$, see Fig.~\ref{fig:sketch}.

As shown in Appendix \ref{derivation}, the quantum master equation for the two-particle state $\rho$ conditioned on the homodyne detection of the scattered light (called the {\it conditional} state) reads \cite{jacobs2014}
\begin{align}\label{mastereq}
     d\rho =& -\frac{i}{\hbar}[H,\rho]dt - \sum_{s=\pm} \left(\frac{D_{\rm g}}{\hbar^2}+\frac{1}{8L^2}\right)[x_s,[x_s,\rho]]dt \nonumber\\ &+  \sum_{r=\text{in,out}}\frac{\sqrt{\eta_r}}{2L} \sum_{s=\pm} \lbrace x_s - \langle x_s \rangle,\rho \rbrace dW_{rs}.
\end{align}
Here,  we included the impact of residual gas collisions with diffusion constant $D_{\rm g} = \gamma_{\text{g}}m k_{\rm B} T_{\rm g}$, where $\gamma_{\rm g}$ denotes the gas damping constant and $T_{\rm g}$ the gas temperature.

The deterministic part of equation (\ref{mastereq}) describes the coherent time evolution of the system induced by the Hamiltonian (\ref{hamiltonian}) as well as the momentum diffusion due to gas collisions and photon recoil. The stochastic term, which is non-linear in the particle state, accounts for the fact that measuring the scattered fields drives the two-particle quantum state into a product of Gaussian states in the sum and difference modes, conditioned on the measurement results (\ref{outcome}). Averaging Eq.~(\ref{mastereq}) over all possible measurement outcomes (and setting $F_{\pm}=0$) yields the master equation for a particle diffusing due to recoil heating and gas collisions \cite{rudolph2021,seberson2020}. Note that the stochastic term effectively reduces the recoil heating and decoherence since a fraction $\eta_r$ of scattered photons is not irretrievably lost.

\section{Force-gradient sensing at the standard quantum limit}We now follow the cold-damping scheme \cite{tebbenjohanns2019,tebbenjohanns2021} and filter the measurement outcomes $dy_{\text{in},s}$ with a feedback filter function $H_s(t)$, that serves as an approximate differentiator \cite{wilson2015,tebbenjohanns2019,tebbenjohanns2021} to mimic linear-velocity damping. Denoting the tunable feedback damping rates by $\gamma_s$, the feedback forces can be written as an It\^{o} stochastic integral
\begin{align}\label{colddampingforce}
    F_s(t) = -m\gamma_s \int_{-\infty}^{\infty}dy_{\text{in},s}(t') H_s(t-t').
\end{align}
Thus, both the position information and the measurement noise are filtered and fed back onto the particle motion. Especially for large feedback rates, this may impact the achievable steady-state energy and the noise floor of force-gradient measurements \cite{tebbenjohanns2019}.

\begin{figure}[t]
	\centering
	\includegraphics[width=1\linewidth]{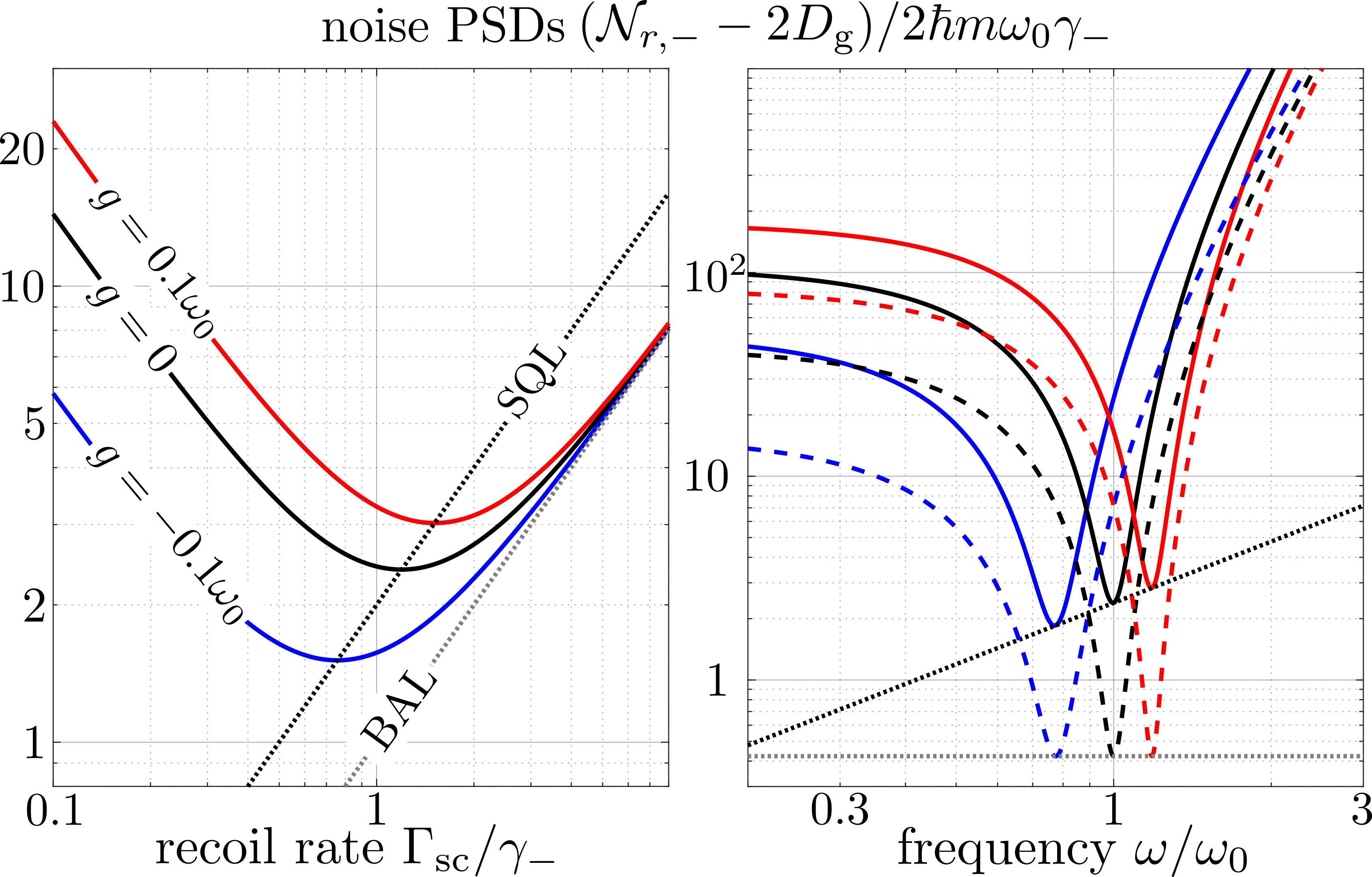}
	\caption{Normalised noise PSDs \eqref{signal} of the difference mode for different coupling constants $g$. Left: $\mathcal{N}_{\text{out},-}$ as function of the recoil heating rate $\Gamma_{\rm sc}$ on resonance ($\omega=\omega_-$). Right, solid: $\mathcal{N}_{\text{out},-}$ as function of the frequency $\omega$ for the fixed recoil rate $\Gamma_{\rm sc}/\gamma_- = \omega_-\sqrt{1/\eta_{\rm in} + 1/\eta_{\rm out}}/4\omega_0$, chosen to minimise the noise on resonance. Right, dashed: The same for $\mathcal{N}_{\text{in},-}$ but with the choice $\Gamma_{\rm sc}/\gamma_- = 1/4\sqrt{\eta_{\rm in}}$. The damping rate is always $\gamma_- =\omega_0/10$ and the in- and out-of-loop efficiencies are $\eta_{\rm in} = 0.35$ and $\eta_{\rm out}=0.05$, respectively. The black solid curve ($g=0$) is also the noise PSD of the sum mode. The black dotted lines connect the minima of all possible $\mathcal{N}_{\text{out},-}$ curves and therefore mark the standard quantum limit (SQL) at resonance. The grey dotted lines describe back-action-limited (BAL) detection at resonance, connecting all dashed lines on the right.}
	\label{fig:Fplot}
\end{figure}

We are now in the position to compute the detected in-loop and out-of-loop power spectral densities (PSD) of the mechanical normal modes. In Appendix \ref{conditionedequations} and \ref{measuredsignal} we show that
\begin{align}\label{eq:psd}
    S_{rs}[\omega] = \frac{|\chi_s[\omega]|^2}{2\pi m^2} \left( 2\pi S_{K_s K_s}[\omega] + \mathcal{N}_{rs}[\omega] \right),
\end{align}
with the mechanical susceptibilities $\chi_s[\omega]=(\omega_s^2-\omega^2 + \gamma_s\sqrt{2\pi}H_s[\omega])^{-1}$ and the noise spectra
\begin{align}\label{signal}
     \mathcal{N}_{rs}[\omega] = & 2D_{\rm g} + \frac{\hbar^2}{4L^2}  \nonumber\\
     &+m^2 L^2\Biggl[\frac{|\chi_{s}[\omega]|^{-2}}{\eta_r}\mp 2\pi\frac{\gamma_s^2}{\eta_{\rm in}} |H_s[\omega]|^2 \Biggr] 
\end{align}
where the negative (positive) sign applies to the in-loop (out-of-loop) PSD.

The minimal force $K_s$ detectable with the PSDs \eqref{eq:psd} is determined by the noise spectra \eqref{signal}. Here, the first two terms describe noise due to gas collisions and measurement back-action (shot noise), while the second line describes imprecision noise of the measurement and noise from feeding the in-loop measurement noise back onto the particle motion. 

At resonance, the in-loop signal is only back-action limited ($D_{\rm g}=0$ in the following) due to squashing of the feedback noise \cite{tebbenjohanns2019}, irrespective of $\gamma_s$. In contrast, the minimal force detectable with the out-of-loop signal at resonance is obtained by choosing $\gamma_s$ as small as technically feasible and minimising with respect to the measurement accuracy. This yields $L^2 \propto 1/\gamma_s$ and
\begin{align}\label{SQL}
    \mathcal{N}_{\text{out},s}[\omega_s]\Bigr \vert_{\rm SQL} = \hbar m\omega_0\gamma_s e^{-2\kappa_s} \sqrt{\frac{1}{\eta_{\rm in}}+\frac{1}{\eta_{\rm out}}},
\end{align}
with the (squeezing) parameter $\kappa_s = \log(\omega_s/\omega_0)/2$. As $\eta_{\rm in} + \eta_{\rm out} \leq 1$, the minimum detectable force is bounded from  below by $2\hbar m \omega_s\gamma_s$, which is reached for $\eta_{\rm in}=\eta_{\rm out}=1/2$. Figure~\ref{fig:Fplot} shows that the SQL of gradient force sensing can drop below the corresponding SQL of two non-interacting particles \cite{li2022}, as given by $g = 0$. Measurements below the SQL of uncoupled oscillators can be achieved if the interaction is repulsive,  and therefore for $\omega_- < \omega_0$, implying higher force susceptibilities. If the SQL cannot be achieved at a given $\gamma_s$, the measurement is still back-action limited.

The force sensitivity of the out-of-loop spectra can be estimated for the experimental setting in Ref.\,\cite{magrini2021} by choosing $\gamma_s$ and $L$ such that the SQL is reached, as $10^{-20}\,\text{N}/\sqrt{\text{Hz}}$, which is about one order of magnitude above the back-action-limit. This implies a force gradient sensitivity on the $\text{zepto-Newton}/\mu\text{m}$ scale for particles levitated at micrometer separations \cite{rieser2022}.

\section{Gaussian entanglement of nanoparticles}For sufficiently strong coupling, the setup in Fig.~\ref{fig:sketch} can also be used to prepare and observe stationary Gaussian entanglement between the two particles in the absence of external forces $K_s = 0$. Specifically, the feedback loop drives the two particles into a non-thermal stationary state, which exhibits gaussian entanglement due to the Coulomb interaction, as quantified by the logarithmic negativity \cite{eisert1999} of the unconditional two-particle state $\mathbb{E}[\rho]$. The continuous observation of the particle motion effectively reduces the recoil heating rate significantly below the Coulomb coupling rate, so that entanglement can be generated and observed by continuously driving the two-particle state into a product of gaussians in the sum and difference modes, as quantified in Appendix \ref{conditionedequations}

Introducing dimensionless position and momentum quadratures via $x_s = \sqrt{\hbar/m\omega_0}X_s$ and $p_s = \sqrt{\hbar m\omega_0}P_s$, one can write the elements of the unconditional covariance matrix as a function of the net heating rate $\Gamma = \Gamma_{\rm sc} + \gamma_{\rm g} k_{\rm B}T_{\rm g}/\hbar\omega_0$ and of the effective detection efficiency $\eta_{\rm eff}=\eta_{\rm in}\Gamma_{\rm sc}/\Gamma$, which describes the detectable fraction of the net information leaving the system. Here, we set $\eta_{\rm out} = 0$ for simplicity, which is sufficient for estimating the achievable entanglement as typically $\eta_{\rm in}\gg\eta_{\rm out}$. To maximize the logarithmic negativity, we consider weak measurements, $\Gamma \ll \omega_s$. In this case one can, following the reasoning in Appendix \ref{entanglingmethods}, approximate the non-vanishing elements of the dimensionless covariance matrix as 
\begin{subequations}\label{covar}
    \begin{align}
       &\mathbb{E}[ \langle X_s^2 \rangle] = \frac{\gamma_s}{16\eta_{\rm eff}\Gamma} + \frac{\omega_0^2 \Gamma}{\omega_s^2 \gamma_s}\\
        &\mathbb{E}[ \langle P_s^2 \rangle] \approx \frac{\omega_s^2}{16\eta_{\rm eff}\omega_0^2\Gamma}\gamma_s + \frac{\Gamma}{\gamma_s} + \frac{\Omega_s\gamma_s^2}{16\eta_{\rm eff}\omega_0^2\Gamma}\\
       &\mathbb{E}\left[\frac 1 2 \langle X_s P_s + \text{h.c.} \rangle\right] \approx  \frac{\gamma_s}{4\sqrt{\eta_{\rm eff}}\omega_s},
    \end{align}
\end{subequations}
where we neglect higher orders in $\Gamma/\omega_s$ and introduce the filter bandwidths
\begin{align}\label{bandwidth}
    \Omega_s = \frac 1 \pi \int_{-\infty}^{\infty} d\omega f_s^2[\omega].
\end{align}
They must exceed the spectral width of the mechanical motion $\Omega_s \gg \gamma_s$ to ensure efficient cooling. Physically, the filter bandwidths appear in \eqref{covar} due to the high frequency fluctuations of the particle momenta originating from feeding back the time derivative of the measurement noise onto the particle dynamics. The optimal choice of the feedback rates for entangling the motion will lead to $\gamma_s \propto \Gamma$, so that $\Omega_s \gg \Gamma$, as shown in Appendix \ref{entanglingmethods}. Note that neither the conditional nor the unconditional two-particle state are thermal.

\begin{figure}[t]
	\centering
	\includegraphics[width=1\linewidth]{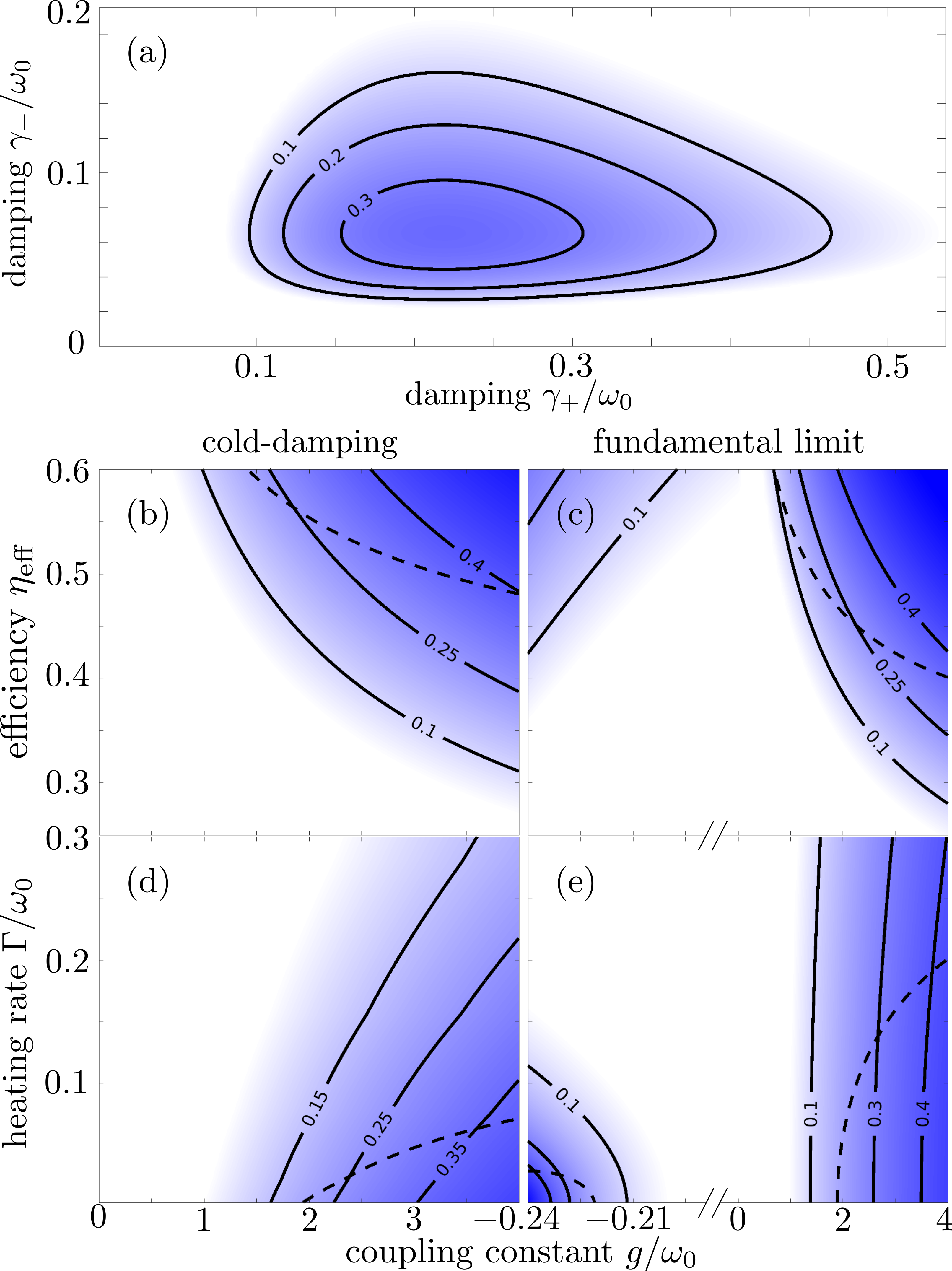}
	\caption{(a) Logarithmic negativity of two interacting, feedback-cooled nanoparticles as a function of the damping constants $\gamma_s$ for an effective detection efficiency of $\eta_{\rm eff}=0.45$, a net heating rate of $\Gamma=0.1\omega_0$ and a coupling constant of $g=4\omega_0$. (b), (d) Logarithmic negativity of the unconditional state as a function of the coupling constant and of the effective detection efficiency or the net heating rate, respectively. The plots are calculated for $\Omega_s=\omega_0$ and at the optimal choice of the damping rates $\gamma_s$. (c), (e) Logarithmic  negativity of the conditional state. The dashed black lines indicate the violation of the Duan-criterion, which is a weaker entanglement criterion than the logarithmic negativity. (b), (c) are calculated for $\Gamma = \omega_0/10$, while (d), (e) assume $\eta_{\rm eff}=0.45$. The unconditional negativity (b), (d) can be realised by cold-damping feedback \cite{tebbenjohanns2021}, while the conditional negativity (c), (e) can be reached by Kalman filtering \cite{magrini2021}.}
	\label{fig:logmain}
\end{figure}

In the limit of weak measurements, $\Gamma \ll \omega_s$, the particles get entangled whenever the detection efficiency exceeds the ratio of the normal mode frequencies, $\eta_{\rm eff} > \omega_</\omega_>$. At finite measurement strengths, the finite filter bandwidth always impairs the creation of entanglement. Using the exact unconditional covariance matrix from Appendix \ref{entanglingmethods}, the logarithmic negativity is depicted in Fig.~\ref{fig:logmain} (a) as function of the two damping constants $\gamma_s$ and in Fig.~\ref{fig:logmain} (b) and (d) as a function of the heating rate, the detection efficiency and the coupling constant, numerically maximised over $\gamma_s$. The plot shows that entanglement can be observed if either the detection efficiency or the coupling constant is sufficiently large. Negative coupling constants can never lead to entanglement for realistic parameters in the cold-damping cooling scheme, since they increase bandwidth-induced fluctuations in Eq.~(\ref{covar}), see Appendix \ref{entanglingmethods}. The logarithmic negativity of the conditional stationary state is shown in Fig.~\ref{fig:logmain} (c) and (e), demonstrating that Kalman filtering can in principle create entanglement even at negative coupling rates \cite{magrini2021}.

\section{Discussion}The presented force sensing and entanglement schemes require the ability to feedback cool both normal modes individually. For short inter-particle distances, the easiest way to achieve this is by applying a homogeneous electric field \cite{wilson2015,tebbenjohanns2019,tebbenjohanns2020,magrini2021,tebbenjohanns2021}, implying that the particle charges must be distinct, $|Q_1|\neq|Q_2|$, to ensure that the feedback acts on both normal modes. Alternatively, all optical cold damping schemes have been demonstrated recently \cite{kamba2022optical,vijadan2022scalable}. To avoid feedback-induced coupling between the normal modes, the bandwidths $\Omega_s$ of the filter functions must be sufficiently narrow  \cite{sommer2019}, $\Omega_s<|\omega_+ - \omega_-|$, and the damping rates sufficiently small, $\gamma_s<|\omega_+ - \omega_-|$.

To show that the presented scheme is feasible, we evaluate it for realistic experimental parameters. A silica sphere ($\chi_{\rm e}=0.8$, $\varrho = 1850\,\text{kg}/\text{m}^3$) with a diameter of 90\,nm, trapped in a laser beam with a wavelength of $2\pi/k=1064\,\text{nm}$, a power of $P=300\,\text{mW}$ and a Rayleigh range of $x_{\rm R}=1.21\,\mu\text{m}$, experiences a trapping frequency of $\omega_0/2\pi= 108\,\text{kHz}$ and a recoil heating rate of $\Gamma_{\rm sc}=\omega_0/10$. In a nitrogen atmosphere with a temperature of $T_{\rm g}=300\,\text{K}$ and a pressure of $10^{-8}\,\text{mbar}$ the gas damping constant is $\gamma_{\rm g}/2\pi=10\,\mu\text{Hz}$ \cite{martinetz2018} and the net heating rate $\Gamma/2\pi = 11.4\,\text{kHz}$. If the in-loop-efficiency is $\eta_{\rm in}=0.4$, which is slightly more ambitious than in \cite{magrini2021}, the effective detection efficiency is $\eta_{\rm eff}=0.38$ (backaction-dominated regime). It follows from Fig.~\ref{fig:logmain} (b) that the coupling constant should be at least about $g>2.5\omega_0$ to achieve entanglement, which can be reached if the particle separation is about $d=2\,\mu\text{m}$ and both particles are charged with about 250 elementary charges, which is in reach of present-day experiments \cite{rieser2022}. 
Note that stable trapping of oppositely charged particles may require compensating their steady-state displacement due to the Coulomb interaction. This can be easily done e.g.\ by applying a homogeneous electrostatic field along the axis connecting the tweezer foci, which might modify the trapping frequency and recoil rate, as shown in Appendix \ref{equilibriumshift}.

To illustrate the force sensitivity and spatial resolution, we note that these parameters would allow for the detection of the force difference 
due to a single electron placed in the plane of the two tweezers, at an angle of $45^\circ$ and a distance of $10\,\mu\text{m}$ from the particles, oscillating with frequency $\omega_-$ and amplitude $1\,\mu\text{m}$  at a Q-factor of 10. The spatial resolution is determined by the inter-particle distance, which can in principle be controlled with an accuracy of 8\,nm \cite{delic2019}.

The discussed schemes can be adapted to Kalman filtering of the mechanical motion \cite{magrini2021}, which may lead to lower motional temperatures and therefore to better sensing precision and stronger entanglement. It also holds the prospect of eliminating the requirement $|\omega_+ - \omega_-|>\Omega_s$, as no additional noise is directly fed into the particle motion.

In conclusion, we presented a scheme to measure and feedback cool two levitated particles interacting via Coulomb forces, and showed how this can give rise to differential force sensing and entanglement of levitated particles.  Recent experimental progress  \cite{magrini2021,tebbenjohanns2021,penny2021,rieser2022,vijadan2022scalable} suggests that it can be realized with state-of-the-art technology. As a major benefit, the presented feedback-based entanglement scheme significantly reduces the requirements on environmental isolation in comparison to pulsed approaches \cite{rudolph2020,weiss2021,cosco2021}. Moreover, our work paves the way for resolving force fields at the zepto-Newton per micrometer scale, with possible  applications for explorations of physics beyond the standard model \cite{moore2021,carney2021,afek2022} and may well find applications in other setups requiring quantum feedback of multiple mechanical modes.

\begin{acknowledgements}
We thank Lorenzo Magrini for helpful discussions. HR, KH and BAS acknowledge funding by the Deutsche Forschungsgemeinschaft (DFG, German Research Foundation)--439339706. UD and MA acknowledge support by the Austrian Science Fund (FWF, Project No. I 5111-N), the European Research Council (ERC 6 CoG QLev4G), by the ERA-NET programme QuantERA under the Grants QuaSeRT and TheBlinQC (via the EC, the Austrian ministries BMDW and BMBWF and research promotion agency FFG), by the European Union’s Horizon 2020 research and innovation programme under Grant No. 863132 (iQLev).
\end{acknowledgements}

\appendix

\section{Derivation of the feedback master equation}\label{derivation} 
This section derives the feedback quantum master equation for a dielectric sub-wavelength particle with volume $V$ and susceptibility tensor $\chi(\Omega)$, depending on  the particle orientation $\Omega$. The particle is illuminated by a monomchromatic optical field of wavenumber $k$, which can always be expressed in the spectral representation \cite{jackson1999}
\begin{align}\label{spectralE}
    \mathbf{E}(\mathbf{r}) = \frac{1}{(2\pi)^{\frac 3 2}} \int d^2\mathbf{n}\, \mathbf{E}(\mathbf{n}) e^{ik\mathbf{n}\cdot \mathbf{r}},
\end{align}
with $\mathbf{n}\cdot \mathbf{E}(\mathbf{n})=0$ due to transversality. For positions far away from the origin, one can expand (\ref{spectralE}) in powers of $1/r$ as
\begin{align}
    \mathbf{E}(r\mathbf{n}) = \frac{1}{(2\pi)^{1/2}ikr}\left[ \mathbf{E}(\mathbf{n})e^{ikr} - \mathbf{E}(-\mathbf{n})e^{-ikr} \right],
\end{align}
neglecting orders of $O(1/r^2)$. Therefore we can identify an outgoing part of the incident wave, proportional to the spectral representation at a given travelling direction $\mathbf{n}$, and an ingoing part, proportional to the spectral representation at $-\mathbf{n}$.

The particle dynamics are determined by the optical potential $V(\mathbf{r},\Omega) = -\varepsilon_0 V \mathbf{E}^*(\mathbf{r})\cdot\chi(\Omega)\mathbf{E}(\mathbf{r})/4$, with $\mathbf{r}$ the particle center of mass, as well as by the scattering Lindblad operators \cite{rudolph2021}
\begin{align}
    L_{\mathbf{n}p}^{\rm sc} = \sqrt{\frac{\varepsilon_0 k^3}{2\hbar}} \frac{V}{4\pi} \mathbf{t}_{\mathbf{n}p}^* \cdot \chi(\Omega)\mathbf{E}(\mathbf{r}) e^{-ik\mathbf{n}\cdot\mathbf{r}}.
\end{align}
Here, $\mathbf{n}$ describes the direction of the scattered photon and $\mathbf{t}_{\mathbf{n}p}$, $p = 1,2$, the two orthogonal polarisation directions. The resulting quantum master equation reads \cite{rudolph2021}
\begin{align}\label{masteralt}
    &\partial_t \rho = \mathcal{L}_0\rho -\frac{i}{\hbar}[V_{\rm opt},\rho]\nonumber\\
    + \int & d^2\mathbf{n} \sum_{p=1,2} \left ( L_{\mathbf{n}p}^{\rm sc} \rho (L_{\mathbf{n}p}^{\rm sc})^\dagger - \frac 1 2 \lbrace (L_{\mathbf{n}p}^{\rm sc})^\dagger L_{\mathbf{n}p}^{\rm sc}, \rho  \rbrace \right ).
\end{align}
The laser-free particle dynamics, described by $\mathcal{L}_0$, includes gas diffusion. 

We now follow the reasoning from \cite{wiseman2009} to derive a master equation conditioned on the outcome of a homodyne measurement on the scattered light. For this we first reformulate equation (\ref{masteralt}) to account for the fact that the total field measured at a detector is given not only by the scattered field but also by the outgoing part of $\mathbf{E}(\mathbf{r})$. Defining the operators
\begin{align}\label{lindblad}
    L_{\mathbf{n}p} = \sqrt{\frac{\varepsilon_0}{4\pi\hbar k^3}}\frac{1}{i} \mathbf{t}_{\mathbf{n}p}^* \cdot \mathbf{E}(\mathbf{n}) + L_{\mathbf{n}p}^{\rm sc},
\end{align}
allows rewriting (\ref{masteralt}) as
\begin{align}\label{masterstart}
    \partial_t \rho = \mathcal{L}_0\rho + \int d^2\mathbf{n} \sum_{p=1,2} \left ( L_{\mathbf{n}p} \rho L_{\mathbf{n}p}^\dagger - \frac 1 2 \lbrace L_{\mathbf{n}p}^\dagger L_{\mathbf{n}p}, \rho  \rbrace\right ).
\end{align}
The laser-induced dynamics are now fully encoded in the new Lindblad operators, which account for the superposition of the scattered field and the outgoing part of the incident field.
Note that
\begin{equation}
    \int d^2\mathbf{n} \sum_{p=1,2} L_{\mathbf{n}p}^\dagger L_{\mathbf{n}p} = \frac{P}{\hbar ck},
\end{equation}
with $P$ the total incident field power.

We next show that the Lindblad operators can be understood as the measurement operators of a detector counting all outgoing photons. In order to model a homodyne measurement at the shot noise limit, all observable photon scattering directions are individually superposed with classical local oscillator fields $\beta_{\mathbf{n}p}$ and then summed over all accessible scattering directions \cite{wiseman2009}. The latter are determined by the solid angle $\Omega_0$ of the light collecting objective. The detector quantum efficiency is denoted as $\eta_0$ and we include dark counts with rate $r$, so that the measurement signal is a Poisson process $dN(t)$ with mean
\begin{align}\label{poisson}
    &\mathbb{E}[dN(t)] = r dt\nonumber\\
    + \eta_0 & \int_{\Omega_0}  d^2\mathbf{n} \sum_{p=1,2} \mathbb{E}[\langle(\beta_{\mathbf{n}p}^* + L_{\mathbf{n}p}^\dagger) (\beta_{\mathbf{n}p} + L_{\mathbf{n}p})\rangle] dt.
\end{align}
From this, we can infer the measurement operators to be $\beta_{\mathbf{n}p}+ L_{\mathbf{n}p}$ \cite{wiseman2009}.

Unravelling the master equation (\ref{masterstart}) with respect to the signal (\ref{poisson}) requires a few steps. We first rewrite (\ref{masterstart}) as
\begin{widetext}
\begin{align}
    &\partial_t \rho = \mathcal{L}_0\rho + \int d^2\mathbf{n} \sum_{p=1,2} \left[ L_{\mathbf{n}p} \rho L_{\mathbf{n}p}^\dagger - \frac 1 2 \lbrace L_{\mathbf{n}p}^\dagger L_{\mathbf{n}p}, \rho  \rbrace \right] - \eta_0\int_{\Omega_0} d^2\mathbf{n} \sum_{p=1,2} \left[ L_{\mathbf{n}p} \rho L_{\mathbf{n}p}^\dagger - \frac 1 2 \lbrace L_{\mathbf{n}p}^\dagger L_{\mathbf{n}p}, \rho  \rbrace \right] \nonumber \\
    & + \eta_0 \int_{\Omega_0} d^2\mathbf{n} \sum_{p=1,2} \left[ (\beta_{\mathbf{n}p} + L_{\mathbf{n}p}) \rho (\beta_{\mathbf{n}p}^* + L_{\mathbf{n}p}^\dagger) - \frac 1 2 \lbrace (\beta_{\mathbf{n}p}^* + L_{\mathbf{n}p}^\dagger) (\beta_{\mathbf{n}p} + L_{\mathbf{n}p}), \rho  \rbrace \right] - \frac{\eta_0}{2} \int_{\Omega_0} d^2 \mathbf{n} \sum_{p = 1,2} \left[ \beta_{\mathbf{n}p}^* L_{\mathbf{n}p} - \beta_{\mathbf{n}p} L_{\mathbf{n}p}^\dagger , \rho \right],
\end{align}
by using that the last term in the first line cancels the second line. Next, the first term in the second line is stochastically unravelled by using (\ref{poisson}),
\begin{align}\label{masterpoisson}
    &d\rho = \mathcal{L}_0\rho dt + \int d^2\mathbf{n} \sum_{p=1,2}\left[ L_{\mathbf{n}p} \rho L_{\mathbf{n}p}^\dagger - \frac 1 2 \lbrace L_{\mathbf{n}p}^\dagger L_{\mathbf{n}p}, \rho  \rbrace \right] dt - \eta_0\int_{\Omega_0} d^2\mathbf{n} \sum_{p=1,2} \left[ L_{\mathbf{n}p} \rho L_{\mathbf{n}p}^\dagger - \frac 1 2 \lbrace L_{\mathbf{n}p}^\dagger L_{\mathbf{n}p}, \rho  \rbrace \right] dt \nonumber \\
    &+ \eta_0 \int_{\Omega_0} d^2\mathbf{n} \sum_{p=1,2} \left[ \langle(\beta_{\mathbf{n}p}^* + L_{\mathbf{n}p}^\dagger) (\beta_{\mathbf{n}p} + L_{\mathbf{n}p})\rangle \rho - \frac 1 2 \lbrace (\beta_{\mathbf{n}p}+ L_{\mathbf{n}p})^\dagger (\beta_{\mathbf{n}p} + L_{\mathbf{n}p}), \rho  \rbrace \right] dt \nonumber \\
    &+\left[\frac{r\rho + \eta_0\int_{\Omega_0} d^2\mathbf{n} \sum_{p=1,2} (\beta_{\mathbf{n}p} + L_{\mathbf{n}p}) \rho (\beta_{\mathbf{n}p}^* + L_{\mathbf{n}p}^\dagger)}{r + \eta_0\int_{\Omega_0} d^2\mathbf{n}\sum_{p=1,2}\langle(\beta_{\mathbf{n}p}^* + L_{\mathbf{n}p}^\dagger) (\beta_{\mathbf{n}p} + L_{\mathbf{n}p})\rangle} - \rho \right] dN(t) - \frac{\eta_0}{2} \int_{\Omega_0} d^2 \mathbf{n} \sum_{p = 1,2} \left[\beta_{\mathbf{n}p}^* L_{\mathbf{n}p} - \beta_{\mathbf{n}p} L_{\mathbf{n}p}^\dagger , \rho \right] dt.
\end{align}
This equation describes the collapse of the quantum state due to the detection or not-detection of a photon. A detection event ($dN = 1$) transforms the state into a mixture of the pre-detection state (if the photon was a dark count) and a superposition state consisting of the pre-detection state (if the photon was from the local oscillator) and the state for a photon leaving the system, averaged over all detection directions $\Omega_0$.

For homodyne detection, the field of the local oscillator is much greater than the field to be detected. In the formal limit $\beta_{\mathbf{n}p}\rightarrow\infty$ the number of photons arriving at the detector in a finite time interval tends to infinity, while the state transformation of a single detection event vanishes. The usual approach to describe this situation slices the time axis into intervals of finite duration $\Delta t$, which must fulfill two conditions \cite{wiseman2009}: First, the slices have to be long enough so that the number of photons arriving at the detector during this period clearly exceeds unity. Second, the slices have to be sufficiently short so that the quantum state changes only weakly during $\Delta t$. The number of photons $\Delta N(t)$ detected during $\Delta t$ is then approximately given by a Gaussian distribution, whose variance equals its mean (\ref{poisson}). One can thus write \cite{wiseman2009}
\begin{align}
    \Delta N(t) \approx \left[r + \eta_0 \int_{\Omega_0} d^2\mathbf{n} \sum_{p=1,2} \langle(\beta_{\mathbf{n}p}^* + L_{\mathbf{n}p}^\dagger) (\beta_{\mathbf{n}p} + L_{\mathbf{n}p})\rangle\right]\Delta t + \sqrt{r + \eta_0 \int_{\Omega_0} d^2\mathbf{n} \sum_{p=1,2} \langle(\beta_{\mathbf{n}p}^* + L_{\mathbf{n}p}^\dagger) (\beta_{\mathbf{n}p} + L_{\mathbf{n}p})\rangle} \Delta W(t)
\end{align}
where $\Delta W(t)$ is a Gaussian distributed random number with zero mean and variance $\Delta t$.

Since $\Delta t$ is much smaller than the evolution time of the master equation, we can replace $dN$ in equation (\ref{masterpoisson}) by $\Delta N$. Keeping only terms to leading order in $\beta_{\mathbf{n}p}^{-1}$, performing the limit $\Delta t \rightarrow dt$, and splitting the Lindblad operators according to equation (\ref{lindblad}) yields
\begin{align}\label{generalfeedback}
    &d\rho = \mathcal{L}_0\rho dt - \frac{i}{\hbar}[V_{\rm opt}(\mathbf{r}),\rho]dt + \int d^2\mathbf{n} \sum_{p=1,2}\left[ L_{\mathbf{n}p}^{\rm sc} \rho (L_{\mathbf{n}p}^{\rm sc})^\dagger - \frac 1 2 \lbrace (L_{\mathbf{n}p}^{\rm sc})^\dagger L_{\mathbf{n}p}^{\rm sc}, \rho  \rbrace \right] dt \nonumber \\
    & + \frac{\eta_0}{\sqrt{r + \eta_0\int_{\Omega_0} d^2\mathbf{n} \sum_{p=1,2}|\beta_{\mathbf{n}p}|^2}} \left[ \int_{\Omega_0} d^2\mathbf{n} \sum_{p=1,2} \beta_{\mathbf{n}p}^* L_{\mathbf{n}p}^{\rm sc}\rho + \beta_{\mathbf{n}p} \rho (L_{\mathbf{n}p}^{\rm sc})^\dagger - \langle \beta_{\mathbf{n}p}^* L_{\mathbf{n}p}^{\rm sc} + \beta_{\mathbf{n}p} (L_{\mathbf{n}p}^{\rm sc})^\dagger \rangle \rho \right] dW(t),
\end{align}
with Wiener increment $dW(t)$. This quantum master equation describes the motion of an arbitrarily shaped nanoparticle subject to continuous homodyning. The homodyne photon flux $I_{\rm hom}$ is obtained by substracting the constant contributions due to the local oscillator, the dark counts, and the incident field $\mathbf{E}$ from the photon count,
\begin{align}
    I_{\rm hom}(t) dt = \eta_0 \int_{\Omega_0} d^2 \mathbf{n}\sum_{p=1,2} \langle \beta_{\mathbf{n}p}^* L_{\mathbf{n}p}^{\rm sc} + \beta_{\mathbf{n}p} (L_{\mathbf{n}p}^{\rm sc})^\dagger \rangle dt + \sqrt{r + \eta_0\int_{\Omega_0} d^2\mathbf{n} \sum_{p=1,2}|\beta_{\mathbf{n}p}|^2}\, dW(t).
\end{align}
The signal is therefore proportional to the scattering field of the particle and is fluctuating due to the photon shot noise of the local oscillator and the dark counts.

As a final step, we now use that the particle is spherical, $\chi(\Omega)=\chi_{\rm e}\mathbb{1} = 3(\varepsilon_{\rm r}-1)\mathbb{1}/(\varepsilon_{\rm r}+2)$, and that the incident field is a tweezer trap of the form
\begin{align}
    \mathbf{E}(\mathbf{r}) = \frac{E_0}{1+ix/x_{\rm R}} \exp\left[-\frac{y^2 + z^2}{w^2 \left(1+i x/x_{\rm R}\right)}\right] e^{i(kx+\varphi_{\rm t})} \mathbf{e}_z,
\end{align}
with waist $w$, Rayleigh range $x_{\rm R}=kw^2/2$, polarisation $\mathbf{e}_z$, amplitude $E_0$, and tweezer phase $\varphi_{\rm t}$. Linearizing Eq.~(\ref{generalfeedback}) around the tweezer focus and tracing out orientational degrees of freedom shows that the deterministic part of the linearised master equation consists of a harmonic trap for all center-of-mass degrees of freedom and Rayleigh scattering diffusion for all components of $\mathbf{r}$ \cite{rudolph2021}. The stochastic term in (\ref{generalfeedback}) can be treated by expanding $L_{\mathbf{n}p}^{\rm sc}$ to the first order in the particle coordinates
\begin{align}\label{Lapprox}
    L_{\mathbf{n}p}^{\rm sc} \approx \sqrt{\frac{\varepsilon_0 k^3}{2\hbar}} \frac{V\chi_{\rm e} E_0}{4\pi} \mathbf{t}_{\mathbf{n}p}^* \cdot \mathbf{e}_z  \left\lbrace1 + i\left[\left(k-\frac{1}{x_{\rm R}}  \right)x - k\mathbf{n}\cdot\mathbf{r}\right]\right\rbrace e^{i\varphi_{\rm t}}.
\end{align}
The constant term cancels and thus does not appear in equation (\ref{generalfeedback}). 

Finally, we assume that the detector solid angle $\Omega_0$ and the local oscillator $\beta_{\mathbf{n}p}$ are chosen such that the measured signal is approximately independent of $y$ and $z$. Tracing out $y$ and $z$ from the master equation and choosing $\mathcal{L}_0$ as the unitary evolution due to the kinetic energy plus a contribution due to gas scattering with diffusion constant $D_{\rm g}$ finally leads to the feedback master equation for the coordinate $x$ along the optical axis,
\begin{align}\label{masterpre}
    d\rho = -\frac{i}{\hbar}[H,\rho]dt - \left(\frac{D_{\rm g}}{\hbar^2}+\frac{1}{8L^2}\right) [x,[x,\rho]]dt + \frac{\sqrt{\eta}}{2L} \left( e^{i\varphi}x\rho + e^{-i\varphi}\rho x - 2\cos\varphi\langle x \rangle \right) dW.
\end{align}
Here, we defined the detection phase
\begin{align}
    \varphi = \varphi_{\rm t} + \arg\left[ \int_{\Omega_0}d^2\mathbf{n}\sum_{p=1,2}i (\mathbf{t}_{\mathbf{n}p}^* \cdot \mathbf{e}_z)\beta_{\mathbf{n}p}^* \left( k-\frac{1}{x_{\rm R}} - k \mathbf{n}\cdot\mathbf{e}_x \right) \right]
\end{align}
and the Hamiltonian
\begin{align}
    H = \frac{p^2}{2m} + \frac{m\omega_0^2}{2}x^2 - F(t)x,
\end{align}
with external force $F(t)$ and trapping frequency $\omega_0 = \sqrt{\chi_{\rm e} k P/\pi c\varrho x_{\text{R}}^3}$. We define the detection efficiency as
\begin{align}
        \eta = \frac{\eta_0^2\left|\int_{\Omega_0}d^2\mathbf{n}\sum_{p=1,2} (\mathbf{t}_{\mathbf{n}p}^*\cdot \mathbf{e}_z)\beta_{\mathbf{n}p}^* \left( k-\frac{1}{x_{\rm R}} - kn_x \right) \right|^2}{\left[\int d^2\mathbf{n}\sum_{p=1,2}|\mathbf{t}_{\mathbf{n}p}\cdot \mathbf{e}_z|^2\left( k-\frac{1}{x_{\rm R}} - kn_x \right)^2\right]\left[r + \eta_0\int_{\Omega_0}d^2\mathbf{n} \sum_{p=1,2} |\beta_{\mathbf{n}p}|^2\right]},
\end{align}
which, following Ref.~\cite{magrini2021}, can be understood as a product of several efficiencies, $\eta = \eta_0 \eta_{\rm inel} \eta_{\rm ov} \eta_{\rm dc}$. Specifically, $\eta_{\rm inel}$ denotes the fraction of inelastically scattered photons arriving at the detector
\begin{align}
    \eta_{\rm inel} = \frac{\int_{\Omega_0}d^2\mathbf{n}\sum_{p=1,2}|\mathbf{t}_{\mathbf{n}p}\cdot \mathbf{e}_z|^2\left( k-\frac{1}{x_{\rm R}} - kn_x \right)^2}{\int d^2\mathbf{n}\sum_{p=1,2}|\mathbf{t}_{\mathbf{n}p}\cdot \mathbf{e}_z|^2\left( k-\frac{1}{x_{\rm R}} - kn_x \right)^2},
\end{align}
while $\eta_{\rm ov}$ describes the mode overlap of the detectable inelastically scattered photons with the local oscillator
\begin{align}
    \eta_{\rm ov} = \frac{\left|\int_{\Omega_0}d^2\mathbf{n}\sum_{p=1,2} (\mathbf{t}_{\mathbf{n}p}^*\cdot \mathbf{e}_z)\beta_{\mathbf{n}p}^* \left( k-\frac{1}{x_{\rm R}} - kn_x \right) \right|^2}{\left[\int_{\Omega_0}d^2\mathbf{n}\sum_{p=1,2}|\mathbf{t}_{\mathbf{n}p}\cdot \mathbf{e}_z|^2\left( k-\frac{1}{x_{\rm R}} - kn_x \right)^2\right]\left[\int_{\Omega_0}d^2\mathbf{n} \sum_{p=1,2} |\beta_{\mathbf{n}p}|^2\right]}.
\end{align}
Likewise, $\eta_{\rm dc}$ is the mean probability for a detected photon to originate from the local oscillator and not from the dark counts
\begin{align}
    \eta_{\rm dc} = \frac{\eta_0\int_{\Omega_0}d^2\mathbf{n} \sum_{p=1,2} |\beta_{\mathbf{n}p}|^2}{r + \eta_0 \int_{\Omega_0}d^2\mathbf{n} \sum_{p=1,2} |\beta_{\mathbf{n}p}|^2}.
\end{align}
We note that most of the weight of $\eta_{\rm inel}$, and therefore most of the particle information, is encoded in the backscattered light, for which  $n_x$ is negative. 

The homodyne photon flux can be written as $I_{\rm hom}(t)dt = I_1 dt + I_2 dy(t)$ with the increment $dy = \langle x \rangle dt \cos\varphi + LdW/\sqrt{\eta}$ and 
\begin{subequations}
    \begin{align}
        I_1 &= \eta_0\int_{\Omega_0}d^2\mathbf{n} \sum_{p = 1,2} \sqrt{\frac{\varepsilon_0 k^3}{2\hbar}}\frac{V\chi_{\rm e}E_0}{4\pi}\beta_{\mathbf{n}p}^* \mathbf{t}_{\mathbf{n}p}^* \cdot \mathbf{e}_z e^{i\varphi_{\rm t}} + \text{c.c.} \\
        I_2 & = \frac{\sqrt{\eta}}{L} \sqrt{r + \eta_0\int_{\Omega_0} d^2\mathbf{n} \sum_{p=1,2}|\beta_{\mathbf{n}p}|^2}.
    \end{align}
\end{subequations}
Substracting the constant photon flux and renormalizing the homodyne signal with $I_2$ thus measures the particle position through $dy(t)$.

Extending this derivation to the case of two particles and two far-detuned tweezers with two individual detectors and transforming to mechanical normal modes, leads us to the feedback master equation (\ref{mastereq}) for $\varphi = 0$. The master equation (\ref{masterpre}) can also be obtained by performing an infinite series of infinitely weak Gaussian measurements on the position operator \cite{jacobs2014}; it is a well-known description applicable to several contemporary experiments \cite{magrini2021,tebbenjohanns2021}.

\section{Equilibrium positions in the presence of Coulomb attraction}\label{equilibriumshift}

The electrostatic interaction not only couples the motion of the two particles along all coordinates, but also displaces their equilibrium positions along their connecting axis ($y$). As the motion along the beam polarisation ($z$) does not influence the particle motion along the optical axis ($x$), we ignore the former in the following considerations.

The potential energy of the $y$- and $x$-coordinates of both particles is given by the sum of the optical potentials of both particles in their respective beams (contributions of the other tweezer can be neglected for all parameters considered in our work) plus the electrostatic interaction. Additionally, a constant and homogeneous electrostatic field $E_{\rm c}$ can be applied along the $y$-axis. The coordinates $\mathbf{r}_{j}=(x_j,y_j,z_j)$ of particle $j=1,2$ refer to the respective beam focus. Then, the potential reads
\begin{align}
    V(x_1,x_2,y_1,y_2) = -\frac{\varepsilon_0 \chi_{\rm e}V}{4} |\mathbf{E}(\mathbf{r}_1)|^2 -\frac{\varepsilon_0 \chi_{\rm e}V}{4} |\mathbf{E}(\mathbf{r}_2)|^2 + \frac{Q_1 Q_2}{4\pi\varepsilon_0 |d \mathbf{e}_y + \mathbf{r}_2 - \mathbf{r}_1|} - Q_1 E_{\rm c} y_1 - Q_2 E_{\rm c} y_2.
\end{align}
For repulsive interaction, $Q_1 Q_2 >0$, we choose $E_{\rm c}=0$. As the motion along the optical axis is only stable for relatively low coupling rates, $g/\omega_0 >-1/4$, the potential can safely be expanded around the two tweezer foci. For attractive interaction and larger coupling rates, however, the constant displacement of the equilibrium positions along $y$ may change the effective trapping frequencies and recoil rates, and may even destabilise the traps. This may be compensated by choosing $E_{\rm c}$ as
\begin{align}
    E_{\rm c} = \frac{Q_1 Q_2}{(Q_1-Q_2)2\pi\varepsilon_0 d^2}.
\end{align}
The resulting potential energy of both particles exhibits extrema at $x_{0,1} = x_{0,2} = 0$ and $y_{0,1} = y_{0,2}$, provided a solution $y_{0,1}$ of the following transcendental equation exists,
\begin{align}
    \frac{y_{0,1}}{w} \exp\left( -2\frac{y_{0,1}^2}{w^2} \right) = - \frac{w d}{x_{\rm R}^2}\left(\frac{g}{\omega_0}\right)_{y=0} \frac{Q_1 + Q_2}{Q_1-Q_2},
\end{align}
where $(g/\omega_0)_{y=0}$ is the ratio of the bare coupling rate and trapping frequency as introduced in the main text. A solution exists if the right-hand side is smaller than the maximum of the left-hand side, $1/2e^{1/2}$, which implies $y_{0,1} < w/2$. The effective trapping frequencies along the optical axis are then given by
\begin{align}
    \omega_0^2 = \frac{\chi_{\rm e} k P}{\pi c\varrho x_{\text{R}}^3} \left(1-2\frac{y_{0,1}^2}{w^2}\right)\exp\left( -2\frac{y_{0,1}^2}{w^2}  \right),
\end{align}
and are thus always reduced in comparison to the bare trapping frequency., Likewise, the recoil heating rate is reduced by a factor of $\exp\left( -2 y_{0,1}^2/w^2\right)$. The coupling rate remains unchanged, apart from its dependence on $\omega_0$. We note that for an equal number of absolute charges on the particles, $|Q_1|=|Q_2|$, the displacement along $y$ can be fully canceled. Equal charges would however prevent feedback cooling of the sum mode via electric fields and thus require optical cold damping cooling \cite{vijadan2022scalable,kamba2022optical}.

The trapping frequency of the center of mass of the particles along $y$ is given by
\begin{align}
    \omega_{y,+}^2 = \frac{4 \chi_{\rm e} P}{\pi c\varrho w^4} \left(1-4\frac{y_{01}^2}{w^2}\right)\exp\left( -2\frac{y_{01}^2}{w^2}  \right),
\end{align}
and the frequency of the difference mode by $\omega_{y,-}^2 = \omega_{y,+}^2 - 4g\omega_0$. Note that the trapping of the difference mode along $y$ can become unstable for sufficiently strong attractive interaction $(\omega_{y,+}^2<4g\omega_0)$, even if $y_{0,1}$ exists. This can however be circumvented with feedback, by superimposing $E_{\rm c}$ with an additional field proportional to $y_2-y_1$ to add an additional restoring force.

\section{Moment equations of motion and stationary conditional covariance}\label{conditionedequations}
From Eq.~(\ref{mastereq}) we can derive the equations of motion for the moments of $x_\pm$ and $p_\pm$. Denoting the conditional covariance of operators $A$ and $B$ as $C_{AB} = \frac 1 2 \langle AB + BA \rangle - \langle A\rangle \langle B \rangle$ and the variance of $A$ as $V_A = C_{AA}$, the first moments can be shown to evolve as
\begin{subequations}\label{firstmomeq}
    \begin{align}
        d\langle x_\pm \rangle &= \frac{\langle p_\pm \rangle}{m}dt + \frac{\sqrt{\eta}}{L}V_{x_\pm} dW_{\pm} + \frac{\sqrt{\eta}}{L}C_{x_+ x_-} dW_{\mp} \\
        d\langle p_\pm \rangle &= -m\omega_\pm^2 \langle x_\pm \rangle dt + F_\pm (t) + K_\pm(t) + \frac{\sqrt{\eta}}{L}C_{x_\pm p_\pm} dW_{\pm} + \frac{\sqrt{\eta}}{L}C_{x_\mp p_\pm} dW_{\mp}.
    \end{align}
\end{subequations}
Here, $\eta=\eta_{\rm in} + \eta_{\rm out}$ and $dW_\pm = (\sqrt{\eta_{\rm in}}dW_{\text{in},\pm} + \sqrt{\eta_{\rm out}}dW_{\text{out},\pm})/\sqrt{\eta}$.

Likewise, the dynamics of the (co-)variances are given by the following set of equations of motion.
\begin{subequations}\label{secmomeq}
    \begin{align}
        dV_{x_\pm} =& \frac{2C_{x_\pm p_\pm}}{m} dt - \frac{\eta}{L^2}V_{x_\pm}^2 dt - \frac{\eta}{L^2} C_{x_+ x_-}^2 dt + \frac{\sqrt{\eta}}{L}(\langle x_\pm^3 \rangle - 3 V_{x_\pm}\langle x_\pm \rangle - \langle x_\pm \rangle^3)dW_\pm \nonumber\\
        &+ \frac{\sqrt{\eta}}{L}(\langle x_\pm^2 x_\mp \rangle - V_{x_\pm}\langle x_\mp \rangle - 2C_{x_+ x_-}\langle x_\pm \rangle - \langle x_\pm \rangle^2)dW_\mp \\
        dC_{x_+ x_-} =& \frac{C_{x_+ p_-} + C_{x_- p_+}}{m}dt - \frac{\eta}{L^2}(V_{x_+} + V_{x_-}) C_{x_+ x_-}dt \nonumber\\
        &+ \frac{\sqrt{\eta}}{L}(\langle x_+^2 x_- \rangle - 2C_{x_+ x_-}\langle x_+ \rangle - V_{x_+} \langle x_- \rangle - \langle x_+ \rangle^2 \langle x_- \rangle)dW_+ \nonumber\\
        &+ \frac{\sqrt{\eta}}{L}(\langle x_-^2 x_+ \rangle - 2C_{x_+ x_-}\langle x_- \rangle - V_{x_-} \langle x_+ \rangle - \langle x_- \rangle^2 \langle x_+ \rangle)dW_- \\
        dC_{x_\pm p_\pm} =& \frac{V_{p_\pm}}{m}dt - m\omega_\pm^2 V_{x_\pm}dt - \frac{\eta}{L^2} V_{x_\pm} C_{x_\pm p_\pm} dt - \frac{\eta}{L^2}C_{x_+ x_-} C_{x_\mp p_\pm} dt \nonumber\\
        &+ \frac{\sqrt{\eta}}{L}(\langle x_\pm p_\pm x_\pm \rangle - 2C_{x_\pm p_\pm}\langle x_\pm \rangle - V_{x_\pm}\langle p_\pm \rangle - \langle x_\pm \rangle^2 \langle p_\pm \rangle)dW_\pm \nonumber\\
        &+ \frac{\sqrt{\eta}}{L}\left(\frac 1 2 \langle x_\pm p_\pm x_\mp + x_\mp p_\pm x_\pm \rangle - C_{x_\pm p_\pm}\langle x_\mp \rangle - C_{x_+ x_-}\langle p_\pm \rangle - C_{x_\mp p_\pm}\langle x_\pm \rangle - \langle x_+ \rangle \langle x_- \rangle \langle p_\pm \rangle\right)dW_\mp \\
        dC_{x_\pm p_\mp} =& \frac{C_{p_+ p_-}}{m}dt - m\omega_\mp^2 C_{x_+ x_-}dt - \frac{\eta}{L^2} V_{x_\pm}C_{x_\pm p_\mp} dt - \frac{\eta}{L^2} C_{x_+ x_-}C_{x_\mp p_\pm} dt \nonumber\\
        &+ \frac{\sqrt{\eta}}{L}(\langle x_\pm^2 p_\mp \rangle - 2C_{x_\pm p_\mp}\langle x_\pm \rangle - V_{x_\pm}^2 \langle p_\mp \rangle - \langle x_\pm \rangle^2 \langle p_\mp \rangle )dW_\pm \nonumber\\
        &+ \frac{\sqrt{\eta}}{L}\left( \frac 1 2 \langle x_\mp p_\mp x_\pm + x_\pm p_\mp x_\mp \rangle - C_{x_\pm p_\mp}\langle x_\mp \rangle - C_{x_\mp p_\mp} \langle x_\pm \rangle - C_{x_+ x_-}\langle p_\mp \rangle - \langle x_+ \rangle\langle x_- \rangle \langle p_\mp \rangle \right)dW_\mp \\
        dV_{p_\pm} =& -2m\omega_\pm^2 C_{x_\pm p_\pm}dt + \left(\frac{\hbar^2}{4L^2} + 2D_{\rm g} \right)dt - \frac{\eta}{L^2}C_{x_\pm p_\pm}^2 dt - \frac{\eta}{L^2} C_{x_\mp p_\pm}^2 dt \nonumber\\
        &+ \frac{\sqrt{\eta}}{L}(\langle p_\pm x_\pm p_\pm \rangle - V_{p_\pm}\langle x_\pm \rangle - 2C_{x_\pm p_\pm} \langle p_\pm \rangle - \langle p_\pm \rangle^2 \langle x_\pm \rangle )dW_\pm \nonumber\\
        &+ \frac{\sqrt{\eta}}{L}(\langle p_\pm^2 x_\mp \rangle - V_{p_\pm}\langle x_\mp \rangle - 2C_{x_\mp p_\pm}\langle p_\pm \rangle - \langle p_\pm \rangle^2 \langle x_\mp \rangle )dW_\mp \\
        dC_{p_+ p_-} =& -m(\omega_+^2 C_{x_+ p_-} + \omega_-^2 C_{x_- p_+})dt - \frac{\eta}{L^2} C_{x_+ p_+}C_{x_+ p_-}dt - \frac{\eta}{L^2} C_{x_- p_-}C_{x_- p_+}dt \nonumber\\
        &+ \frac{\sqrt{\eta}}{L}\left( \frac 1 2 \langle p_- x_+ p_+ + p_+ x_+ p_- \rangle - C_{p_+ p_-}\langle x_+ \rangle - C_{x_+ p_+}\langle p_- \rangle - C_{x_+ p_-} \langle p_+ \rangle - \langle p_+ \rangle \langle p_- \rangle \langle x_+ \rangle \right)dW_+ \nonumber\\
        &+ \frac{\sqrt{\eta}}{L}\left( \frac 1 2 \langle p_+ x_- p_- + p_- x_- p_+ \rangle - C_{p_+ p_-}\langle x_- \rangle - C_{x_- p_-}\langle p_+ \rangle - C_{x_- p_+} \langle p_- \rangle - \langle p_+ \rangle \langle p_- \rangle \langle x_- \rangle \right)dW_-.
    \end{align}
\end{subequations}
In general, neither the first and second moments nor the subsystems $s=\pm$ decouple from each other since all equations of motion are non-linearly coupled. Moreover, the equations of motion are not closed because third moments appear in the stochastic part of the (co-)variances. While all moments are driven by all measurement noises $dW_{r\pm}$,  the (co-)variance equations of motion do not depend on the feedback and external forces $F_\pm$ and $K_\pm$, which has also been noticed for a single particle \cite{bowen2015,magrini2021,tebbenjohanns2021}.

The equations (\ref{firstmomeq}) and (\ref{secmomeq}) cannot be solved in general. However, the problem simplifies greatly for Gaussian states, because they remain Gaussian for all times under equation (\ref{masterpre}). To see this, consider an infinitesimal time step, which can be written as $\rho(t+dt)\propto\int_{-\infty}^\infty du \exp[-u^2/2] \mathcal{W}(u)\rho(t)\mathcal{W}^\dagger(u)$, where
\begin{align}
    \mathcal{W}(u) = \exp\left[ -\frac{i}{\hbar}H_u dt -\eta \frac{(x\cos\varphi - dy(t)/dt )^2}{4L^2} dt \right],
\end{align}
with
\begin{align}
    H_u = H + \frac{\hbar\eta}{4L^2}\cos\varphi\sin\varphi\, x^2 - \hbar\left( \frac{\sqrt{\eta}}{2L}\sin\varphi \frac{dW(t)}{dt} + u\sqrt{\frac{2D_{\rm g}}{\hbar^2} + \frac{1-\eta}{4L^2}} dt^{-1/2} + \frac{\eta}{2L^2}\cos\varphi\sin\varphi \, \langle x \rangle \right)x.
\end{align}
As long as $H$ is at most quadratic in all position and momentum operators, the master equation thus conserves the initial state's gaussianity.

Gaussian states fulfill the following relations for their third moments
\begin{align}\label{eq:B4}
    \langle :u v w: \rangle = C_{uv}\langle w \rangle + C_{uw} \langle v \rangle + C_{vw} \langle u \rangle + \langle u \rangle \langle v \rangle \langle w \rangle,
\end{align}
with $u$, $v$ and $w$ position or momentum operators and $: uvw :$ denoting their Weyl ordering. This can be shown with the help of the Wigner representation of a Gaussian state by using that the Weyl symbol of a Weyl-ordered operator product is the product of the individual Weyl symbols. The above relation then follows from a classical calculation, using that the Wigner function is a Gaussian.

Using Eq.~\eqref{eq:B4} we see that all stochastic terms in the dynamics of the second moments (\ref{secmomeq}) vanish, giving rise to a set of coupled and nonlinear, but closed and deterministic differential equations. After a transient time, the (co-)variances settle at the stationary solutions  
\begin{subequations}
    \begin{align}
        V_{x_s} =& \frac{\sqrt{2}L^2\omega_s}{\eta}\zeta_s \\
        C_{x_s p_s} =& \frac{mL^2\omega_s^2}{\eta} \zeta_s^2 \\
        V_{p_s} = & m^2\omega_s^2 V_{x_s} (1+\zeta_s^2),
    \end{align}
\end{subequations}
and $C_{x_+ x_-}=C_{p_+ p_-}=C_{x_\pm p_\mp} = 0$, where $s\in\lbrace +,- \rbrace$ and
\begin{align}
    \zeta_s^2 = \sqrt{1 + \frac{\eta(\hbar^2 + 8D_{\rm g}L^2)}{4m^2 L^4 \omega_s^4}} - 1.
\end{align}
This implies that the conditional sum and difference modes become uncorrelated. The steady state of the conditional covariance matrix is only a property of the measurement setting and not of the external forces or of the applied feedback. 

\section{Measurement signals}\label{measuredsignal}

Inserting the stationary covariance into Eq.~(\ref{firstmomeq}) demonstrates that the sum and difference motion is only driven by the sum and difference noises $dW_{r\pm}$, respectively. 

Inserting the cold-damping feedback force (\ref{colddampingforce}), the equations of motion for the first moments (\ref{firstmomeq}) can be solved in the frequency domain. Denoting all frequency-dependent quantities with square brackets, we define the Fourier transform by $A[\omega]=\int_{-\infty}^{\infty} dt\,e^{i\omega t} A(t)/\sqrt{2\pi}$. The stationary first moments are then given by
\begin{subequations}\label{firstmomentssolution}
    \begin{align}
        \langle x_s \rangle [\omega] &= \chi_s[\omega] \left[ \left( \frac{C_{x_s p_s}}{m} - i\omega V_{x_s} \right)\frac{\sqrt{\eta}}{L}\xi_{s}[\omega] - \frac{\gamma_s L}{\sqrt{\eta_{\rm in}}}\sqrt{2\pi}H_s[\omega]\xi_{\text{in},s}[\omega] + \frac{K_s[\omega]}{m} \right] ,\\
        \langle p_s \rangle [\omega] &= -m\chi_s[\omega] \left[ \left(i\omega\frac{C_{x_s p_s}}{m} + \omega_s^2 V_{x_s} + \gamma_s\sqrt{2\pi} H_s[\omega] V_{x_s} \right) \frac{\sqrt{\eta}}{L}\xi_s[\omega] -i\omega\frac{\gamma_s L}{\sqrt{\eta_{\rm in}}}\sqrt{2\pi}H_s[\omega] \xi_{\text{in},s} + i\omega \frac{K_s[\omega]}{m}\right],
    \end{align}
\end{subequations}
with the mechanical susceptibilities $\chi_s[\omega] = (\omega_s^2 - \omega^2 + \sqrt{2\pi}\gamma_s H_s [\omega])^{-1}$ and the white noises $\xi[\omega]$, as labeled by different subscripts. The latter are the Fourier transforms of the respective  $dW/dt$, so that $\mathbb{E}[\xi[\omega]] = 0$ and $\mathbb{E}[\xi[\omega']^* \xi[\omega]] = \delta(\omega-\omega')$. From these relations one can directly calculate all power spectral densities $S_{AB}[\omega] = \int_{-\infty}^{\infty}d\omega' \mathbb{E}[\langle A \rangle [\omega]^* \langle B \rangle[\omega']]/2\pi$.

At this point we note that the feedback filter function is causal, $H_s(t<0)=0$, and real-valued, implying that the Fourier transform $H_s[\omega]$ has non-vanishing real and imaginary parts. Specifically, the reality of $H_s(t)$ implies $H_s[\omega] = H_s^*[-\omega]$, so that we can write $H_s[\omega] = (g_s[\omega]-i\omega f_s[\omega])/\sqrt{2\pi}$ with real and symmetric functions $g_s[\omega]$ and $f_s[\omega]$. For the filter function to generate cold damping with rate $\gamma_s$, we require $g_s[\omega_s]=0$ and $f_s[\omega_s]=1$, with both functions approximately constant for a spectral width of at least $\gamma_s$ around $\omega_s$. We note that in realistic situations, the filter $H_s(t)$ is always compact, so that $H_s[\omega]$ decays as $1/\omega$ for large $\omega$, implying that $f_s[\omega]$ decays as $1/\omega^2$.

First, we calculate the PSDs of the in-loop measurement signals, $dy_{\text{in},s}(t)/dt$. They are given by
\begin{align}
    S_{\text{in},s}[\omega] = \frac{L^2}{2\pi \eta_{\rm in}}+ \frac{|\chi_s[\omega]|^2}{2\pi}\Biggl[ \frac{\hbar^2}{4m^2 L^2} + \frac{2D_{\rm g}}{m^2} - 2\pi\frac{L^2}{\eta_{\rm in}}\gamma_s^2|H_s[\omega]|^2 - 2(\omega_s^2 - \omega^2)\frac{L^2}{\eta_{\rm in}}\gamma_s g_s[\omega] \Biggr] + |\chi_s[\omega]|^2 \frac{S_{K_s K_s}[\omega]}{m^2},
\end{align}
showing noise squashing \cite{tebbenjohanns2019} for sufficienctly large damping constants. Neglecting $g_s[\omega]$ yields Eq.\,\eqref{eq:psd} and \eqref{signal} in the main text. Likewise, the PSDs of the out-of-loop signals $dy_{\text{out},s}(t)/dt$ are also given by Eq.~(\ref{eq:psd}) and \eqref{signal}. 
\end{widetext}

\section{Entangling two particles by Coulomb interaction}\label{entanglingmethods}
For Gaussian quantum states the amount of entanglement is encoded in the covariance matrix of the two particles. We set $K_s = 0$ and $\eta_{\rm out}=0$ in the following, so that $\eta = \eta_{\rm in}$. To get the unconditional covariance matrix, one may calculate the PSDs $S_{x_s x_s}$, $S_{x_s p_s}$ and $S_{p_s p_s}$ from Eqs.~(\ref{firstmomentssolution}) and integrate them over all frequencies yielding $\mathbb{E}[\langle x_s \rangle^2]$, $\mathbb{E}[\langle x_s \rangle\langle p_s \rangle]$ and $\mathbb{E}[\langle p_s \rangle^2]$.

In order to evaluate these integrals, we will restrict ourselves to cases where $f_s[\omega]$ and $g_s[\omega]$ change weakly around $\omega_s$ in an interval  larger than $\gamma_s$. This applies if the filter $H_s[\omega]$ is much broader than the mechanical PSD. Then, we can set $H_s[\omega]\approx H_s[\omega_s]$ for integrating over $S_{x_s x_s}$ and $S_{x_s p_s}$. Note that the ensemble average correlations between sum and difference modes are zero. When integrating over $S_{p_s p_s}$, however, the finite cutoff of $H_s[\omega]$ plays a role due to high-frequency fluctuations emerging from feeding back the filtered measurement noise into the oscillators. This manifests itself in a term of the form $f_s[\omega]^2 \omega^4 |\chi_s[\omega]|^2$ in the integrand, where $f_s[\omega]$ determines the convergence of the integral, rather than $|\chi_s[\omega]|^2$. Keeping the exact form of $g_s$ and $f_s$ only where they are needed to achieve convergence yields
\begin{subequations}
    \begin{align}
        \mathbb{E}[\langle x_s \rangle^2] =& \frac{L^2}{2\eta}\gamma_s + \left( \frac{\hbar^2}{8L^2} + D_{\rm g} \right)\frac{1}{m^2\omega_s^2 \gamma_s} - V_{x_s} \\
        \mathbb{E}[\langle p_s \rangle^2] =& m^2\omega_s^2 \Biggl[ \mathbb{E}[\langle x_s \rangle^2] + V_{x_s}\nonumber\\
        +& \frac{L^2}{\eta}\left( \zeta_s^2\gamma_s + \frac{\Omega_s}{2\omega_s^2}\gamma_s^2 \right) \Biggr] - V_{p_s} \\
        \mathbb{E}[\langle x_s \rangle \langle p_s \rangle] =& \frac{mL^2\omega_s}{\sqrt{2}\eta}\zeta_s\gamma_s - C_{x_s p_s},
    \end{align}
\end{subequations}
with the filter bandwidths $\Omega_s$. The latter satisfy $\Omega_s \gg \gamma_s$ by construction (and would diverge if $H_s$ acted as an exact derivative). The integral is well-defined in all realistic cases due to the $1/\omega^2$-decay of the integrand. Finally adding the conditional covariance matrix to these results leads us to the unconditional covariance matrix
\begin{subequations}
    \begin{align}
        &\mathbb{E}[\langle x_s^2 \rangle] = \frac{L^2}{2\eta}\gamma_s + \left( \frac{\hbar^2}{8L^2} + D_{\rm g} \right)\frac{1}{m^2\omega_s^2 \gamma_s} \\
        &\mathbb{E}[\langle p_s^2 \rangle] =m^2\omega_s^2 \left[ \mathbb{E}[\langle x_s^2 \rangle] + \frac{L^2}{\eta}\left( \zeta_s^2\gamma_s + \frac{\Omega_s}{2\omega_s^2}\gamma_s^2 \right) \right] \\
        &\mathbb{E}\left[\frac 1 2\langle x_s p_s + p_s x_s \rangle\right] =\frac{mL^2\omega_s}{\sqrt{2}\eta}\zeta_s\gamma_s.
    \end{align}
\end{subequations}
All other second moments vanish.

Note that the equipartition theorem does not hold in general because the stationary state is not thermal. This is due to the fact that the positions, but not the momenta, of the particles are be measured, leading to an additional momentum uncertainty manifesting in the $\zeta_s$-term in the momentum variances. This term vanishes for sufficiently large measurement uncertainties $L$ \cite{tebbenjohanns2021}. In addition, the high-frequency fluctuations in the position measurement signal lead to a contribution of the feedback bandwidth to the momentum variances, which also violates the equipartition theorem. 

\begin{figure}[t]
	\centering
	\includegraphics[width=1\linewidth]{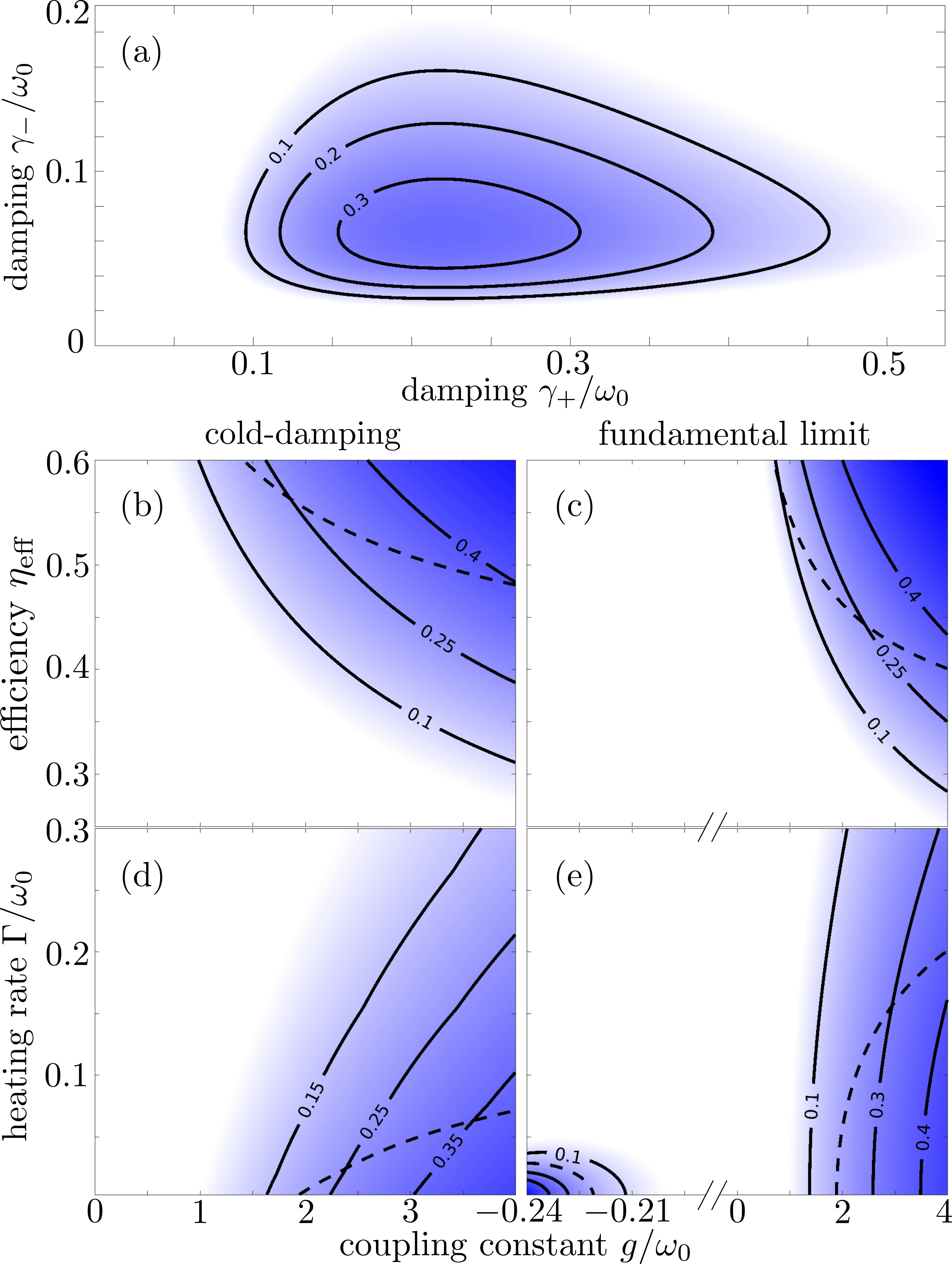}
	\caption{(a) Logarithmic negativity, using the approximation $ c_{1,2}^2 \approx \mathbb{E}[ \langle X_\pm^2 \rangle]\mathbb{E}[ \langle P_\mp^2 \rangle]$ for the symplectic eigenvalues, of two interacting, feedback-cooled nanoparticles as a function of the damping constants $\gamma_s$ for an effective detection efficiency of $\eta_{\rm eff}=0.45$, a net heating rate of $\Gamma=\omega_0/10$ and a coupling constant of $g=4\omega_0$. (b), (d) Approximated logarithmic negativity of the unconditional state as a function of the coupling constant and of the effective detection efficiency or the net heating rate, respectively. The plots are calculated for $\Omega_s=\omega_0$ and at the optimal choice of the damping rates $\gamma_s$. (c), (e) Approximated logarithmic  negativity of the conditional state. The dashed black lines indicate the violation of the Duan-criterion. (b), (c) are calculated for $\Gamma = \omega_0/10$, while (d), (e) assume $\eta_{\rm eff}=0.45$.}
	\label{fig:logappendix}
\end{figure}

To quantify the amount of entanglement present in the two-particle stationary state, we calculate its logarithmic negativity. Recalling that the main text introduces dimensionless position and momentum quadratures via $x_s = \sqrt{\hbar/m\omega_0}X_s$, $p_s = \sqrt{\hbar m\omega_0}P_s$ as well as the net heating rate $\Gamma = \Gamma_{\rm sc}+\gamma_{\rm g}k_{\rm B}T_{\rm g}/\hbar\omega_0$ and the effective detection efficiency $\eta_{\rm eff} = \eta_{\rm in} \Gamma_{\rm sc}/\Gamma$, the elements of the dimensionless conditional covariance matrix can be written as
\begin{subequations}
    \begin{align}
        V_{X_s} &= \frac{\sqrt{\sqrt{\omega_s^4 + 16\eta_{\rm eff}\omega_0^2\Gamma^2}-\omega_s^2}}{4\sqrt{2}\eta_{\rm eff}\Gamma}\\
        V_{P_s} &= \frac{\omega_s^2}{\omega_0^2} V_{X_s} + \frac{32\eta_{\rm eff}^2\Gamma^2}{\omega_0^2}V_{X_s}^3 \\
        C_{X_s P_s} &= \frac{4\eta_{\rm eff}\Gamma}{\omega_0} V_{X_s}^2.
    \end{align}
\end{subequations}
Thus, the purity of the conditional state is $\eta_{\rm eff}$, as estimated from  Eq.~(7) in \cite{paris2003}. The ensemble-averaged second moments (unconditional covariances) are
\begin{subequations}\label{secmoms}
    \begin{align}
       \mathbb{E}[ \langle X_s^2 \rangle] =& \frac{\gamma_s}{16\eta_{\rm eff}\Gamma} + \frac{\omega_0^2 \Gamma}{\omega_s^2 \gamma_s}\\
       \mathbb{E}[ \langle P_s^2 \rangle] =& \frac{2\sqrt{\omega_s^4 + 16\eta_{\rm eff}\omega_0^2\Gamma^2}-\omega_s^2}{16\eta_{\rm eff}\omega_0^2\Gamma}\gamma_s\nonumber\\
       &+ \frac{\Gamma}{\gamma_s} + \frac{\Omega_s\gamma_s^2}{16\eta_{\rm eff}\omega_0^2\Gamma}\\
       \mathbb{E}\Biggl[\frac 1 2 \langle X_s P_s & + P_s X_s \rangle\Biggr] =  \frac{\sqrt{\sqrt{\omega_s^4 + 16\eta_{\rm eff}\omega_0^2\Gamma^2}-\omega_s^2}}{8\sqrt{2}\eta_{\rm eff}\omega_0\Gamma}\gamma_s.
    \end{align}
\end{subequations}
For bipartite Gaussian states the logarithmic negativity can be written as $E_N = \text{max}[0,-\log_2{(2\text{min}[c_1,c_2])}]$, where $c_{1,2}$ are the symplectic eigenvalues of the partially transposed covariance matrix \cite{vidal2002}. In our system, where all correlations between sum and difference mode vanish, the symplectic eigenvalues are
\begin{widetext}
\begin{align}\label{symplectic}
    c_{1,2}^2=&\frac 1 2 \left( V_{X_+}V_{P_-} + V_{X_-}V_{P_+} - 2 C_{X_+ P_+} C_{X_- P_-} \right)\nonumber\\ &\pm \sqrt{\frac 1 4 \left( V_{X_+}V_{P_-} - V_{X_-}V_{P_+} \right)^2 + \left( V_{X_-}C_{X_+ P_+} - V_{X_+}C_{X_- P_-} \right)\left( V_{P_-}C_{X_+ P_+} - V_{P_+}C_{X_- P_-} \right)}.
\end{align}
\end{widetext}
First, we calculate the negativity of the conditional state, which acts as a fundamental limit of optimised cooling schemes \cite{magrini2021,tebbenjohanns2021}. We restrict our discussion to weak measurements, where $\Gamma \ll \omega_0$, for simplicity. Then, in the first non-vanishing order of $\Gamma/\omega_0$, the position variance can be written as
\begin{align}
    V_{X_s} \approx \frac{\omega_0}{2\sqrt{\eta_{\rm eff}}\omega_s} - \frac{\sqrt{\eta_{\rm eff}}\omega_0^3\Gamma^2}{\omega_s^5}.
\end{align}
Inserting it into $V_{P_s}$ and $C_{X_s P_s}$ and calculating the symplectic eigenvalues to the first non-vanishing order in $\Gamma/\omega_0$ yields the conditional minimum symplectic eigenvalue
\begin{align}
    \text{min}(c_1,c_2)\approx \sqrt{\frac{\omega_<}{4\eta_{\rm eff}\omega_>}}\left[ 1 + \frac{\eta_{\rm eff}\Gamma^2}{\omega_0^2} h\left(\frac{\omega_0}{\omega_-}\right) \right],
\end{align}
which must fulfill $\text{min}(c_1,c_2)<1/2$ for the particles to be entangled. Here, we defined $h(s) = |1-s|(s^4 + 2s^3 + 2s + 1)/(1+s)$. The logarithmic negativity is then given by
\begin{align}
    E_N \approx \text{max}\left[0, \frac 1 2 \log_2{\frac{\omega_>\eta_{\rm eff}}{\omega_<}} - \frac{\eta_{\rm eff}\Gamma^2}{\ln{2}\,\omega_0^2} h \left(\frac{\omega_0}{\omega_-}\right) \right].
\end{align}
In the limit $\Gamma/\omega_0 \rightarrow 0$ the correlator $C_{X_s P_s}$ vanishes and the symplectic eigenvalues take the simple form $\sqrt{V_{X_\pm} V_{P_\mp}}$ \cite{ludwig2010}. Then, the conditional state of the particles is entangled if the condition $\eta_{\rm eff}>\omega_</\omega_>$ is fulfilled. In Fig.~\ref{fig:logappendix} (b)-(e) we show the logarithmic negativity as a function of coupling constant, heating rate, and effective detection efficiency if using the approximation $\sqrt{V_{X_\pm} V_{P_\mp}}$ as symplectic eigenvalues. Comparing it to Fig.~\ref{fig:logmain}, we see that the simplified expression predicts weaker entanglement than the exact negativity, but is still a stronger criterion than the Duan inequality.

Second, we will turn to the entanglement of the unconditional state, as achievable with cold-damping feedback. This requires replacing the (co-)variances in Eq.~(\ref{symplectic}) with the unconditional second moments (\ref{secmoms}). Again, we restrict the discussion to $\Gamma\ll\omega_0$ for simplicity. Since the damping constants $\gamma_s$ can be freely chosen we always take them to maximise the entanglement between the two particles. The dependence of the logarithmic negativity on the two damping constants $\gamma_s$, depicted in Fig.~\ref{fig:logmain} (a), shows a strong maximum for specific choices of the $\gamma_s$.

We will see that the optimal choice leads to $\gamma_s \propto \Gamma$. Therefore, considering only the first non-vanishing order in $\Gamma/\omega_0$ leads to Eq.~(\ref{covar}), where the term due to the filter width $\Omega_s$ is always greater than corrections on the order of $\Gamma^2/\omega_0^2$ because $\Omega_s \gg \Gamma$.

We note that the correlators $\mathbb{E}\left[\frac 1 2 \langle X_s P_s + P_s X_s \rangle\right]$ are on the order of $\Gamma/\omega_0$ and appear quadratically in the logarithmic negativity. They can thus be neglected, so that agin the simplified formula $ c_{1,2}^2 \approx \mathbb{E}[ \langle X_\pm^2 \rangle]\mathbb{E}[ \langle P_\mp^2 \rangle]$ for the symplectic eigenvalues can be used, see also Fig.~\ref{fig:logappendix}. Hence, when maximising the negativity we can minimise $\mathbb{E}[\langle X_\pm^2 \rangle]$ and $\mathbb{E}[\langle P_\mp^2 \rangle]$ individually. To minimise the position variance we choose $\gamma_\pm = 4\Gamma\sqrt{\eta_{\rm eff}}\omega_0/\omega_\pm$, while the momentum fluctuations can be minimised to leading order in $\Gamma/\omega_0$ with
\begin{align}\label{optimalgamma}
    \gamma_\mp \approx 4\Gamma\sqrt{\eta_{\rm eff}} \frac{\omega_0}{\omega_\mp} - \frac{16\eta_{\rm eff}\omega_0^2\Omega_\mp\Gamma^2}{\omega_\mp^4}.
\end{align}
These choices lead to
\begin{subequations}
    \begin{align}
        &\mathbb{E}[ \langle X_\pm^2 \rangle] = \frac{\omega_0}{2\sqrt{\eta_{\rm  tot}}\omega_\pm}\\
        &\mathbb{E}[ \langle P_\mp^2 \rangle] \approx \frac{\omega_\mp}{2\sqrt{\eta_{\rm eff}}\omega_0} + \frac{\Omega_\mp\Gamma}{\omega_\mp^2}.
    \end{align}
\end{subequations}
The minimum symplectic eigenvalue then reads
\begin{align}
    \text{min}(c_1,c_2) \approx \sqrt{\frac{\omega_<}{4\eta_{\rm eff}\omega_>}}\left(1 + \sqrt{\eta_{\rm eff}}\frac{\omega_>\Omega_<\Gamma}{\omega_- \omega_<^2} \right),
\end{align}
allowing us to estimate the logarithmic negativity to leading order in $\Gamma/\omega_0$,
\begin{align}\label{negativity}
    E_N &\approx \text{max}\left[0, \frac 1 2 \log_2{\left(\eta_{\rm eff}\frac{\omega_>}{\omega_<}\right)} - \frac{\sqrt{\eta_{\rm eff}}}{\ln{2}}\frac{\omega_>}{\omega_<}\frac{\Omega_<}{\omega_<}\frac{\Gamma}{\omega_-} \right],
\end{align}
where $\omega_{<}=\min(\omega_+,\omega_-)$ and $\omega_{>}=\max(\omega_+,\omega_-)$, with $\Omega_<$ the associated filter bandwidth. In the limit of weak coupling, $|g|\ll \omega_0$, and $\Omega_+ \approx \Omega_-$, the entanglement criterion can be simplified to $|g|>2n_+\omega_0$, where $n_+ = \mathbb{E}[\langle X_+^2 + P_+^2 - 1 \rangle]/2$ is the stationary occupation of the sum mode. The resulting logarithmic negativity is given by $E_N \approx \text{max}[0,|g|/\omega_0 -2n_+]/\ln{2}$, resembling the expression for particles coupled to a cold bath \cite{ludwig2010}. 

In the limit of $\Gamma/\omega_0 \rightarrow 0$, the logarithmic negativities of the conditional and of the unconditional state become identical. The logarithmic negativity when approximating the symplectic eigenvalues with $c_{1,2}^2 \approx \mathbb{E}[\langle X_\pm^2 \rangle] \mathbb{E}[\langle P_\mp^2 \rangle]$ is depicted in Fig.~\ref{fig:logappendix} (a), (b) and (d), where we numerically maximised the approximated negativity (or the violation of the Duan criterion) in (b) and (d) with respect to $\gamma_s$. It demonstrates, together with Fig.~\ref{fig:logmain}, that one needs larger coupling constants and greater efficiencies to violate the Duan criterion than needed to generate entanglement.

\end{document}